\documentclass[prb,onecolumn,12pt]{revtex4}
\usepackage{epsfig}
\usepackage{amssymb}
\usepackage{threeparttable}
\usepackage{graphicx}
\usepackage{natbib}
\usepackage{longtable}
\usepackage{txfonts}

\begin{document}
\title{Coupling of evanescent waves into propagation channels within two-dimensional random waveguides}
\author{Dequan Zhang$^1$, Yuchen Xu$^1$, Ke Xu$^1$, Hao Zhang$^{1,2,*}$, Jing Li$^1$, Heyuan Zhu$^{1}$, Costas Soukoulis$^{2,3}$}
\affiliation{
$^1$Department of Optical Science and Engineering and Key Laboratory of Micro and Nano Photonic Structures (Ministry of Education), Fudan University, Shanghai 200433, China\\
$^2$Department of Physics and Astronomy and Ames Laboratory, Iowa State University, Ames, Iowa 50011, USA\\
$^3$Institute of Electronic Structure and Laser (IESL), FORTH, 71110 Heraklion, Crete, Greece}
\email{zhangh@fudan.edu.cn}

\begin{abstract}
The transformation from evanescent waves to propagation waves is the key mechanism for the realization of some super-resolution imaging methods. By using the recursive Green function and scattering-matrix theory, we investigated in details on the transport of evanescent waves through a random medium and analyzed quantitatively the coupling of evanescent channels to propagation channels. By numerical calculations, we found that the transmission for the incident evanescent channel is determined by both the eigenvalues of the scattering matrix and the coupling strength to the corresponding propagation channels in random medium, and    
the disorder strength of the random medium influences both of them.   
\end{abstract}

\maketitle

\section{Introduction}
In the past two decades, many new principles and schemes to realize an imaging beyond the diffraction limit ($\sim\lambda/2$) have been proposed\cite{zhuang2010,betzig1992,pendry2000,liu2007,liu2007nl,Lerosey2007,Putten2011,Rotter2017}, which generally can be divided into two categories, i.e. super-solution fluorescence microscopy techniques and noninvasive and label-free super-solution microscopy techniques, e.g. atomic force microscopy (AFM), the Scanning Near-Field Optical Microscopy (SNOM) and etc. The latter relies on quantum or purely optical technologies. As for the optical super-resolution microscopy techniques such as SNOM, from the view point of optics, to realize the so-called super-resolved imaging, it is required to deliver and recover the subwavelength information from the object onto the image.

The subwavelength information of an object is carried by evanescent waves, which decay exponentially in space and vanish in the order of wavelength $\lambda$, leading to the diffraction limit in conventional optics. To realize the far-field super-resolution imaging, one approach is to recover the subwavelength information carried by evanescent waves by placing a subwavelength-sized aperture or sharp stylus in the near-field region ($<\lambda$) of the object and converting the evanescent waves into propagating waves, which can be detected in the far-field microscope. The resolution limit of microscopy technologies based on such near-field imaging mechanism\cite{betzig1992,Zenhausern1995}, e.g. SNOM\cite{ASH1972,Hecht2000}, is determined by the size of the aperture or sharp stylus and reaches the resolution of tens of nanometers ($\sim\lambda/4$) recently\cite{Taminiau2007}. 

In addition to the near-field optical microscopy techniques, Pendry \textit{et. al.} proposed the concept of superlens, which consists a thin slab composed of the material with negative permittivity, permeability or both\cite{pendry2000,smith2005,fang2005}. Superlens transmits evanescent waves by the excitation of surface plasmons and allows the recovery of evanescent waves at the image\cite{fang2005}. Since no transformation from evanescent to propagating waves take place, superlens can only produce a near-field super-resolved imaging. To convert evanescent waves to propagating waves and realize far-field super-resolved imaging, the concept of hyperlens is proposed\cite{jacob2006,liu2007}, which is constructed by cylindrical layered metamaterials with a hyperbolic dispersion, allowing the propagation of waves with very large spatial frequency. i.e. evanescent waves in vacuum, which finally recover the subwavelength information at the far-field image.         

Recently an approach for subwavelength imaging based on the transformation of evanescent waves to propagating waves and time-reversal principle was proposed\cite{Lerosey2007}, and the optical configuration composes a microstructured random media in the near-field region of the object to convert evanescent waves to propagating waves, and a far-field ($\sim10\lambda$) time-reversal mirror (TRM) to build the time-reversed wave field, which will finally form a focus at the target. The resolution of such a method for microwaves reaches a value as small as one-thirtieth of a wavelength\cite{Lerosey2007}. Obviously, the transformation mechanism of evanescent waves to propagating waves plays a crucial role in the explanation of such super-resolution imaging. 

However, although such a super-resolution imaging method has been widely used in telecommunication, and much effort has been devoted on the investigation and control of propagation of light through a random medium to realize the subwavelength imaging\cite{Vellekoop2007,Popoff2010,Putten2011}, the conversion of evanescent waves to propagation waves in the random medium is still lack of quantitative analysis.

In this work, we report a quantitative analysis on the coupling of incident evanescent waves to the propagation waves in the random medium by the numerical investigation on the light transport from a two-dimensional homogeneous waveguide through a two-dimensional random waveguide. The energy carried by the evanescent channel is coupled to and then delivered by the propagation channel in the random medium and then coupled to the outgoing channels. The disorder strength of the random medium  will influence the energy distribution within the propagation channels and determine the corresponding channel transmission.  

\section{Numerical Model and Methods}

\begin{figure}
    \centering
    \includegraphics[width=0.6\textwidth]{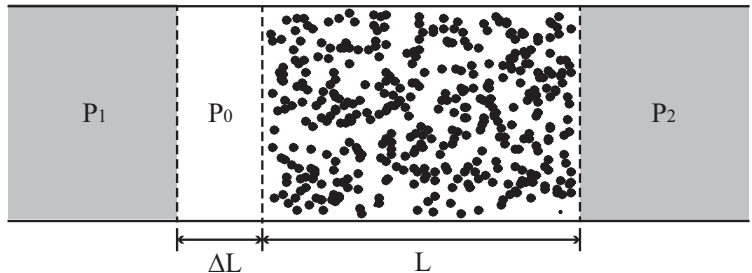}
    \caption{Schematic view of two-dimensional random waveguide.}
    \label{model}
\end{figure}

To simplify the problem, here we consider a simple model composed of a two-dimensional random waveguide (2DRW) attached by two 2D semi-infinite homogeneous waveguides, as shown in Fig. \ref{model}. The width of the 2DRW is $W$, the left/right free waveguides with respective refractive indices of $n_1$ and $n_2$ are denoted by $P_1/P_2$. For simplification, the imaginary parts of the dielectric constants are neglected. The width of the random scatterers $RS$ is $L$, and the distribution of the dielectric function of $RS$ region is described by

\begin{equation}
\epsilon(x,y)=n_0^2[1+\mu(x,y)]
\end{equation}

where $\mu(x,y)$ is distributed in a homogeneous random pattern in the region $[-\sigma.\sigma]$, in which $\sigma$ is the disorder parameter. In order to generate an incident evanescent wave, in the region between $P_1$ and $RS$, a homogeneous waveguide with the size of $\Delta L$ is inserted, and denoted by $P_0$. The refractive index of $P_0$ is $n_0$.

The transverse boundaries of the 2DRW system are perfectly reflective, and a monochromatic and z-polarized light $E(x,y)e^{i\omega t}$ propagates along the x direction, which is governed by the Helmholtz equation,

\begin{equation}
[\nabla^2+k_0^2\epsilon(x,y)]E(x,y)=0
\label{helmholtz}
\end{equation}

where $k_0=\omega/c$ is the light wavevector in vaccum, and $\epsilon(x,y)$ is the spatial distribution of dielectric functions for the 2DRW system. 

Within the semi-infinite homogeneous waveguides $P_{1/2}$, according to the waveguide theory, the propagation modes can be quantized into $N$ transverse modes, which are generally called as ``channels''\cite{Beenakker1997,xu2017}. In the context of the scattering-matrix theory\cite{Beenakker1997,Fisher1981}, the transport of light through a random-scattering region can be described by the scattering matrix $t$, whose elements $t_{\beta\alpha}$ represents the complex field transmission amplitude from the incoming channel $\alpha$ to the outgoing channle $\beta$ and can be calculated by the Fisher-Lee relation\cite{Fisher1981},

\begin{equation}
t_{\beta\alpha}=\sqrt{v_\beta v_\alpha}\int_0^Wdy\int_0^Wdy^\prime \chi_\beta^*(y)G^r(L,y;0,y^\prime)\chi_\alpha(y^\prime)
\label{tba}
\end{equation}

where $v_i$ is the group velocity of the $i^{th}$ channel in the corresponding homogeneous waveguide, $\chi_i(y)$ is the transverse wave function of the $i^{th}$ mode which has the form of standing waves due to the perfect reflections on both transverse boundaries. $G^r(L,y;0,y^\prime)$ is the retarded Green function describing light propagation from the source point $(0,y^\prime)$ to the probe point $(L,y)$, which can be calculated by the recursive Green function (RGF) method\cite{MacKinnon1985,Baranger1991,xu2016,xu2017}.

If the left and right semi-infinite homogeneous waveguides possess an identical width and dielectric constant, the number of propagation modes $N$ they support are identical. The total transmission for an incident channel $\alpha$, i.e. $T_\alpha$, can be calculated by summation over the contribution from all the outgoing channels $\beta$, as follows,

\begin{equation}
T_\alpha=\sum_{\beta=1}^N|t_{\beta\alpha}|^2
\label{ta}
\end{equation}

When the size of $P_0$ is zero, i.e. $\Delta L=0$, since the two homogeneous waveguides in the left and right sides of the $RS$ region support propagation modes with different number, i.e. $N_1$ and $N_2$ ($N_1 \neq N_2$, due to $n_1\neq n_2$), the dimension of the scattering matrix $t$ for light propagating through the $RS$ region is $N_2\times N_1$. The singular-value decomposition of $t$ can be written as,

\begin{equation}
t=U\Lambda V^\dag
\label{ULV}
\end{equation} 
  
where the matrix $U$ is the $N_1\times N_1$ matrix mapping the incident channels in the left waveguide to the eigen channels in the $RS$ region, while the matrix $V$ is the $N_2\times N_2$ matrix mapping the eigen channels in the $RS$ region to the output channels in the right waveguide. 

If $N_1>N_2$, the diagonal matrix $\Lambda$ possesses $N_2$ singular values $\tau_n$ of the scattering matrix $t$, and $\tau_1>\tau_1>..>\tau_{N_2}>0$\cite{mello1988}, which are regarded as propagation channels in the waveguides. In fact, one can obtain $N_1$ eigenvalues by diagonalizing the $N_1\times N_1$ Hermitian matrix $t^\dag t$, however, the values for the modes larger than $N_2$ are equal to zero, i.e. $\tau_{N_2+1},\tau_{N_2+2},..,\tau_{N_1}=0$, which can be regarded as the evanescent channels in the $RS$ region.

\section{Numerical Results}
\subsection{Wave transport within homogeneous waveguides $P_{0/1/2}$}
As we know, within homogeneous waveguide, the number of the propagating modes can be obtained as,
\begin{equation}\label{nwg}
N=\frac{nk_0W}{\pi}
\end{equation}
where $n$ is the refractive index of the waveguide, and $W$ is the width of the waveguide. For the homogeneous waveguides $P_{1/2}$ shown in Fig. \ref{model}, the numbers of the propagating modes are different since $n_1\neq n_2$. Here, we denote the number of the propagating modes in homogeneous waveguides $P_{0/1/2}$ as $N_{0/1/2}$, respectively.

When the refractive index of $P_1$ is larger than that of $P_0$, i.e.$n_1>n_0$, $P_1$ waveguide will support more propagating modes than $P_0$, i.e. $N_1>N_0$, according to Eq. (\ref{nwg}), which means that some higher-order propagating modes in $P_1$ will be totally reflected at the interface between $P_1$ and $P_0$, and emerge as evanescent mdoes in $P_0$, which will propagate within a short distance in the order of wavelength and then vanish.  

For the channel $\alpha$ in the homogeneous waveguide $P_{0/1/2}$, the relation between $k_{\alpha,x}$ and $k_{\alpha,y}$ can be written as,
\begin{equation}
k_{\alpha,x}^2+k_{\alpha,y}^2=(n_ik_0)^2
\end{equation}
where $n_i$ is the refractive index in the $P_i$ region. According to the waveguide theory, the normal component of wavevector $k_{\alpha,y}=\alpha\pi/(N+1)$, and the propagating modes thus can be sorted by the propagation angles $\theta$ determined by $\theta_\alpha=arctan(k_{\alpha,y}/k_{\alpha,x})$.

For simplicity without loss of generality, here we choose the geometric parameters and materials properties of the waveguide model as $W=50/k_0$, $n_0=1.0$, $n_1=1.16$, $n_2=1.0$ and $L=100/k_0$. The widths of the respective waveguide regions are normalized by the light wavevector in vaccum $k_0$. According to Eq. (\ref{nwg}), the numbers of propagating modes within $P_{0/1/2} $ are $N_1=19$ and $N_0=N_2=16$, respectively. The propagating modes $\alpha$ are sorted by the incident angles $\theta_\alpha$. As mentioned above, there are three modes, i.e. modes No. 17-19, propagating within $P_1$ will be totally reflected at the interface between $P_1$ and $P_0$, since $P_0$ can support only 16 propagating modes.

\subsection{Propagating modes through the RS region}
\begin{figure}
    \centering
    \includegraphics[width=0.6\textwidth]{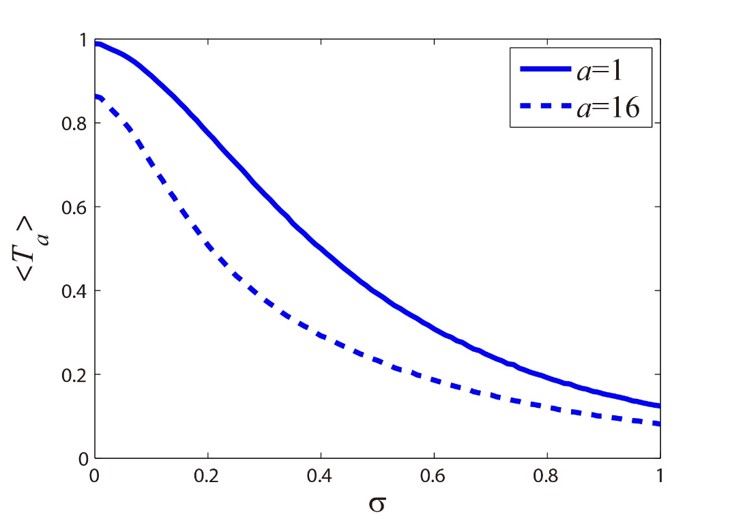}
    \caption{Disorder-dependence of averaging transmission $<T_\alpha>$ for first and last propagating modes, i.e. propagation modes No. 1 and 16.}
    \label{propagation}
\end{figure}

To investigate the influence on the propagation modes from the different levels of random scattering in the $RS$ region, the width of the $P_0$ is set to be zero, i.e. $\Delta L=0$, and then we changed the disorder parameter $\sigma$ in the $RS$ region. By using the RGF method described above, the averaging transmission for different incident channels $\alpha$, i.e. $<T_\alpha>$, is calculated for random system with different disorder parameter $\sigma$, as shown in Fig. \ref{propagation}.

It is clearly shown that, for the propagation modes propagating within this 2DRW, the value of $<T_\alpha>$ decreases when the disorder parameter $\sigma$ in the $RS$ region increases, which is due to the scattering events occur in the $RS$ region.

\subsection{Evanescent modes through the RS region}
\begin{figure}
    \centering
    \includegraphics[width=0.6\textwidth]{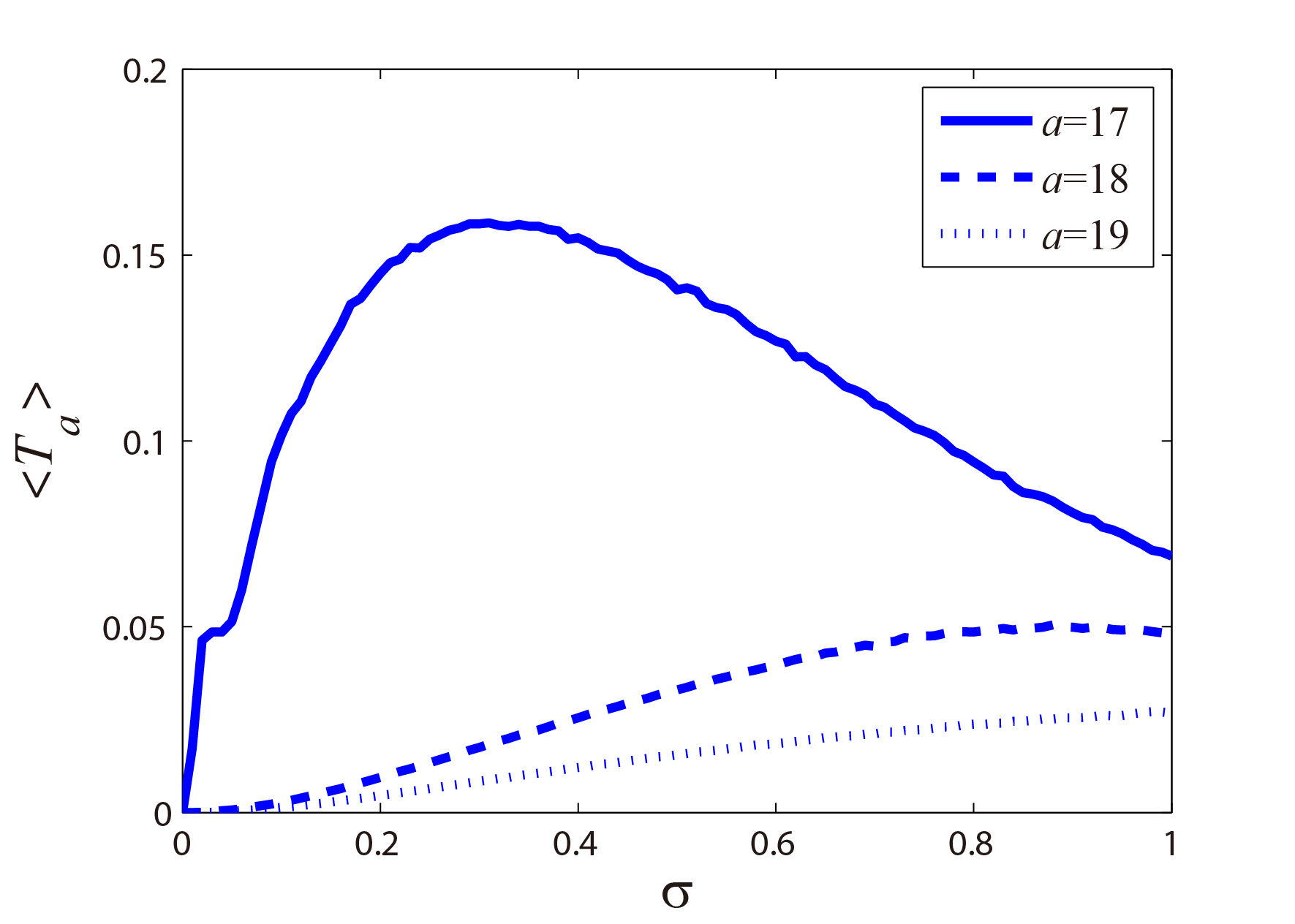}
    \caption{Disorder-dependence of Averaging transmission $<T_\alpha>$ for evanescent modes.}
    \label{sigma}
\end{figure}

For comparison, the behaviors of averaging transmission $<T_\alpha>$ for evanescent modes, i.e. modes No. 17-19, are shown in Fig~\ref{sigma}. When the $RS$ region is free from randomly distributed scatterers, i.e. $\sigma=0$, the values of averaging transmission $<T_\alpha>$ for all evanescent modes are equal to zero, as shown in Fig. \ref{sigma}, which is due to the fact that propagation modes No. 17-19 in the waveguide $P_1$ are totally reflected at the interface between $P_1$ and $RS$ (note that $\Delta L=0$) and the evanescent modes can propagate nontrivially within a short distance $\sim \lambda$. However, when random scatterings indeed occur in the $RS$ region, i.e. $\sigma\neq0$, the behaviors of $<T_\alpha>$ for evanescent modes are quite different from those for propagation modes shown in Fig. \ref{propagation}. For evanescent modes No. 17-18, there are some peaks in the $<T_\alpha>$ curves and No. 19 mode increases when the disorder parameter $\sigma$ increases. Since the amplitude of $<T_\alpha>$ is far from zero, it can come to the conclusion that the evanescent channels within the $RS$ region also carry nontrivial energy.

\begin{figure}
    \centering
    \includegraphics[width=0.6\textwidth]{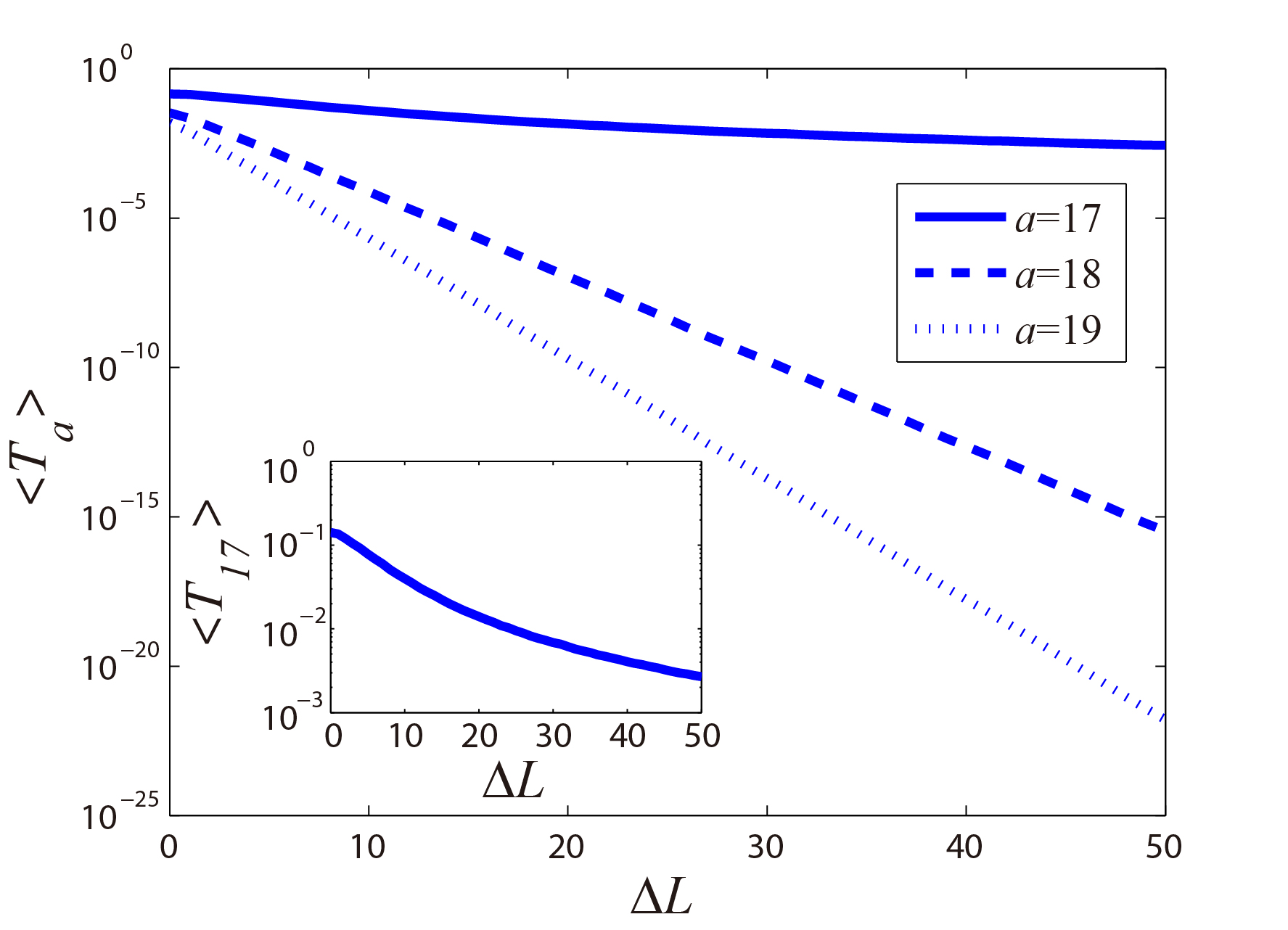}
    \caption{$\Delta L$-dependence averaging transmission $<T_\alpha>$ for the incident modes $\alpha=17-19$.}
    \label{width}
\end{figure}

In order to eliminate the influence from the interface scattering taking place at the interface between the $P_1$ and $RS$ regions, in Fig. \ref{width}, we calculate the $<T_\alpha>$ for the case with an increasing $\Delta L$ and a fixed disorder parameter $\sigma=0.5$. For modes No. 18 and 19, the values of $<T_\alpha>$ decrease exponentially  when the size of $P_0$ increases, which confirms that these two modes are evanescent within the $P_0$ region. The inclination of $\log<T_\alpha(\Delta L)>$ for mode 19 is much larger than that for the mode 18 is due to the fact that $|k_{19,x}|>|k_{18,x}|$ caused by $k_{19,y}<k_{18,y}$.

However, the case for the incident mode 17 whose incident angle is close to the total-reflection angle, is quite different, the value of $<T_\alpha(\Delta L)>$ keeps nontrivial even when the size of $P_0$, i.e. $\Delta L\sim50/k_0$, shown as the inset in Fig. \ref{width}, probably due to the small mismatch between the incident mode 17 and the eigen-mode of the $RS$ region.

\subsection{Averaging transmission $<\tau_n>$ of the RS region}

To further investigate the underlying mechanism of the nontrivial transmission for incident evanescent channels, we set the size of $P_0$ to zero, i.e. $\Delta L=0$, and the disorder parameter $\sigma$ is fixed to 0.5, to calculate the averaging eigen values $\tau_n$ of the scattering matrix $t$ for $RS$ region. As mentioned above, since the number of propagation modes in $P_1$ is larger than that in $P_2$, i.e. $N_1(=19)>N_2(=16)$, the eigen values $\tau_n$ of the scattering matrix $t$ for the $RS$ region possess $N_2$ nonzero values, and the eigen values for modes larger than $N_2$ are all equal to zero, i.e.

\begin{equation}
\tau_{17}=\tau_{18}=\tau_{19}=0
\label{tau1719}
\end{equation}

\begin{figure}
    \centering
    \includegraphics[width=0.6\textwidth]{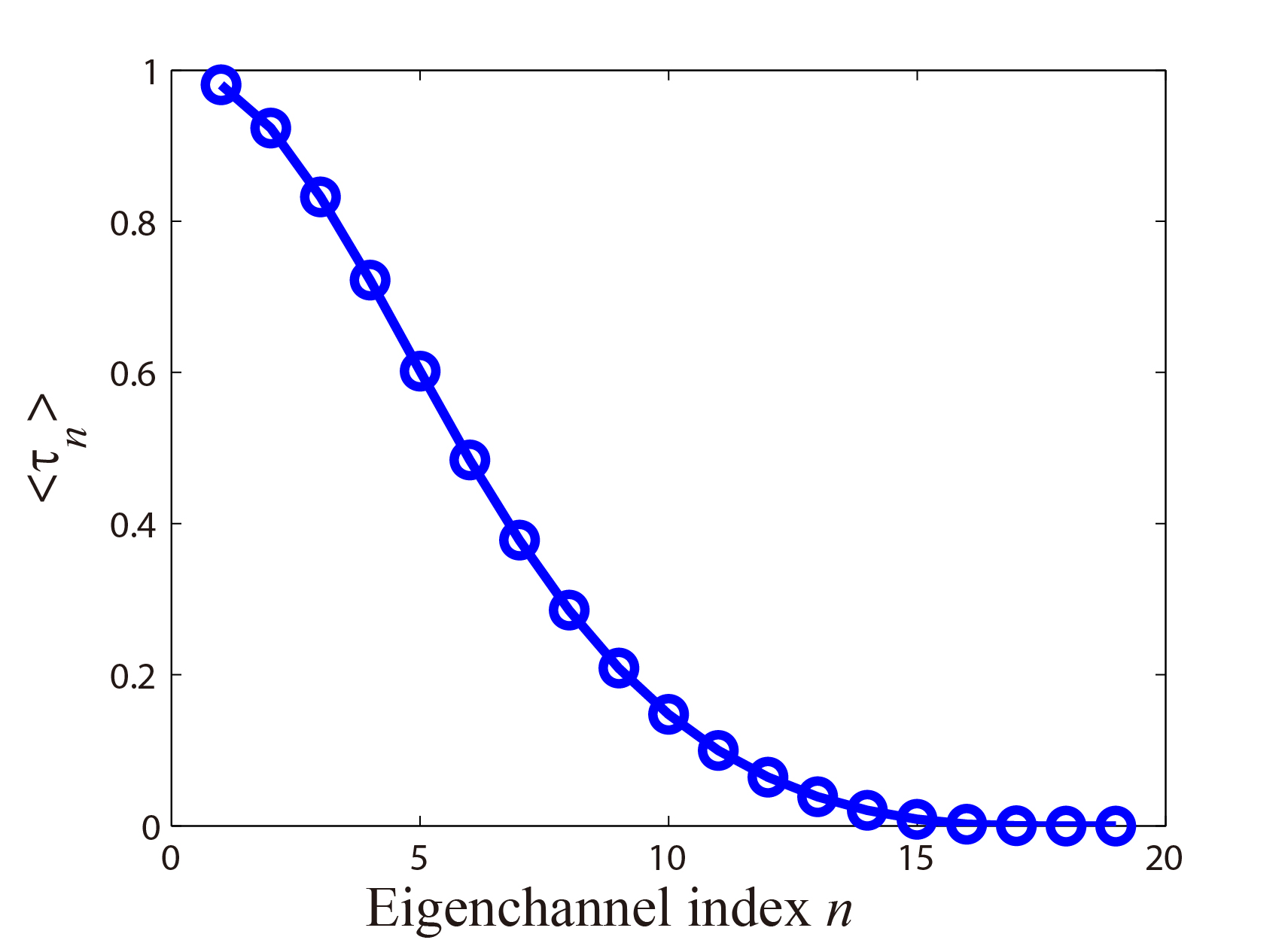}
    \caption{Averaging transmission $<\tau_n>$ for different eigen channels in the $RS$ region.}
    \label{tau}
\end{figure}

By using the RGF method, we calculate the averaging eigen values $<\tau_n>$ of the scattering matrix $t$ for the $RS$ region, which are also the averaging transmission $T_\alpha$ for different incident channels $\alpha$. The calculated results are shown in Fig. \ref{tau}, which also validates the analystical expression for the eigen channels for the $RS$ region, as described by Eq. (\ref{tau1719}).    

\section{Discussions}
From the averaging eigen values $<\tau_n>$ of the scattering matrix $t$ for the $RS$ region, as shown in Fig. \ref{tau}, the evanescent channels of the $RS$ region, i.e. eigenchannels No. 17-19 here, can neither propagate and nor carry nontrivial energy and thus behave like evanescent waves. Thus, the nontrivial transmission for the incident evanescent modes from $P_1$ can not be understood by the coupling of incident energy into the evanescent channels of the $RS$ region. 

According to the scattering matrix method, the coupling coefficient for the incident channel $\alpha$ in $P_1$ waveguide into the eigenchannel $n$ in the $RS$ region is the element of the mapping matrix $V$ in Eq. (\ref{ULV}), i.e. $V_{\alpha n}$, and the coupling strength can be defined as $|V_{\alpha n}|^2$. It should be noted that, as shown in Fig. \ref{tau}, the energy coupled into the evanescent channel of the $RS$ region is equal to zero and contributes nothing to the total transmission. For the incident channel $\alpha$, the transmission $T_\alpha$ can be written as,

\begin{eqnarray}
T_\alpha=&\sum_{\beta}|t_{\beta\alpha}|^2 \nonumber\\ 
=&\sum_{\beta}(\sum_{n^\prime}U_{\beta n^\prime}^*\sqrt{\tau_{ n^\prime}}V_{\alpha n^\prime})(\sum_{n^\prime}U_{\beta n^\prime}\sqrt{\tau_n}V_{\alpha n^\prime}^*) \nonumber\\
=&\sum_n|V_{\alpha n}|^2\tau_n
\label{taeq}
\end{eqnarray}

According to Eq. (\ref{taeq}), the total transmission for an incident channel $\alpha$, i.e. $T_\alpha$, is determined by the weighted summation over the contribution from different eigenchannels of the $RS$ region, and the weighted factor is the coupling strength $|V_{\alpha n}|^2$. Furthermore, no coupling between the eigenchannels of the $RS$ region can be found. Therefore, the total transmission for an incident channel $\alpha$ is determined by both the coupling strength $|V_{\alpha n}|^2$ and eigen values $\tau_n$ of the scattering matrix in $RS$ region, which is the reason why the coupling into evanescent channels in the $RS$ region contributes nothing to the total transmission.
  
\begin{figure}
    \centering
    \includegraphics[width=0.6\textwidth]{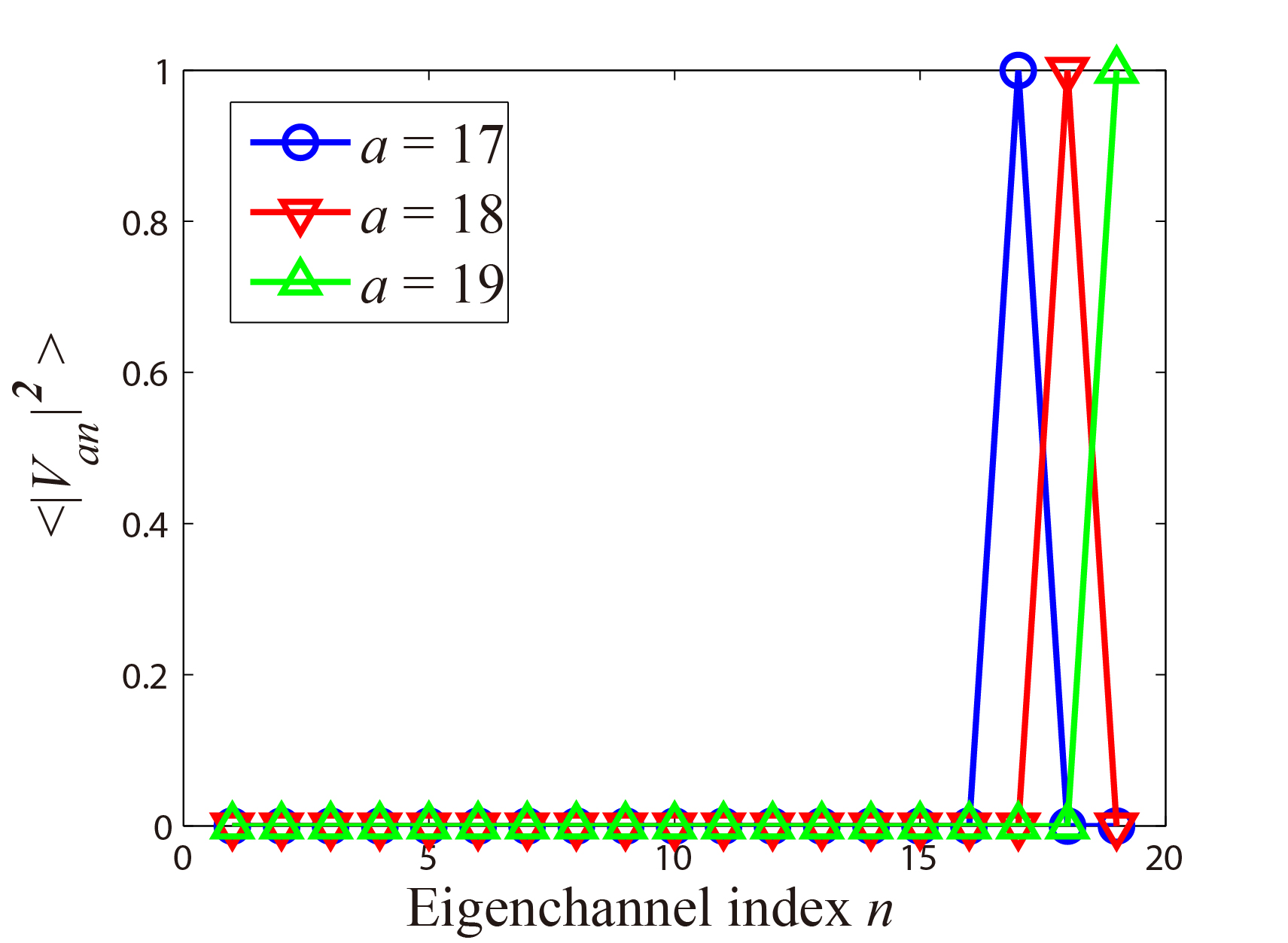}
    \caption{Averaging coupling strength $<|V_{\alpha,n}|^2>$ for different eigen channels.}
    \label{vans0}
\end{figure}

By ensemble averaging the total transmission $T_\alpha$ for an incident channel $\alpha$, the average coupling strength for channel $\alpha$ can be defined as,

\begin{equation}
<|V_{\alpha n}|^2>=\frac{<T_\alpha>}{<\tau_n>}
\end{equation} 

\begin{figure}
    \centering
    \includegraphics[width=0.6\textwidth]{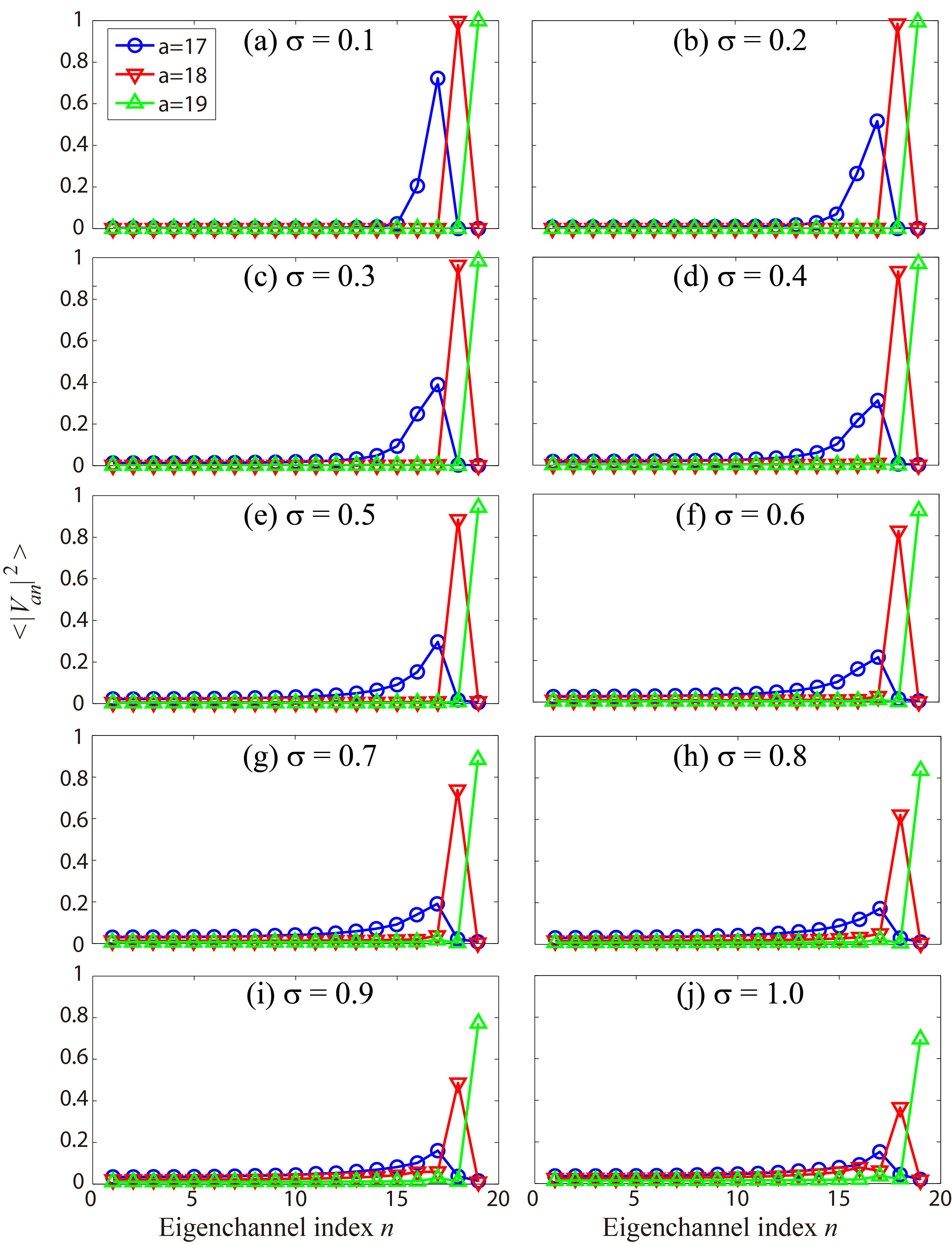}
    \caption{Dependence of the averaging coupling strength $<|V_{\alpha,n}|^2>$ on the disorder parameter $\sigma$ for different evanescent channels.}
    \label{vans}
\end{figure}

For comparison, a series of samples with different disorder parameters $\sigma$ are investigated, and only the evanescent channels are considered, as shown in Figs \ref{vans0}-\ref{vans}. When the $RS$ region is free from random scatterers, i.e. $\sigma=0$, and can be regarded as homogeneous waveguide with $n=n_0$, the calculated averaging coupling strengths $<|V_{\alpha n}|^2>$ for the incident evanescent channels, i.e. ($\alpha=17-19$), have the value of one, as shown in Fig. \ref{vans0}, which means that these three evanescent channels in $P_1$ waveguide are perfectly coupled to the corresponding evanescent channels in the $RS$ region. This is easy to understand by the fact that, under such a condition, regions of $P_1$ and $RS$ are homogeneous waveguides with identical dielectric constant and width.

Since the eigen values for the evanescent channels in the $RS$ region are all equal to zero ($\tau_{17-19}=0$) and carrying no energy, therefore, the corresponding transmissions are equal to zero, i.e. $T_{17-19}=0$. In such situation, total reflections happen at the interface between the $P_1$ and $RS$ regions for these evanescent channels.

When random scatterings are introduced, i.e. $\sigma\neq0$, it has been shown in Figs. \ref{sigma}-\ref{width} that nontrivial transmitted energy emerge from the interface between the $RS$ and $P_2$ for incident evanescent channels. The averaging coupling strength for evanescent channels are shown in Fig. \ref{vans}. 

For all the evanescent channels, it is shown in Fig. \ref{vans} that the maximum coupling strength is the coupling to the corresponding channel with the same index, i.e. $max[<|V_{\alpha n}|^2>]=<|V_{\alpha\alpha}|^2>$, due to the small mismatch between the incident channel and the channel in the $RS$ region with the same channel index.

When the disorder parameter $\sigma$ increases, the coupling from incident evanescent channels in $P_1$ waveguide to the propagation channels in the $RS$ region become nontrivial. Here, taking the case of channel 17 as an example, the averaging coupling strength to propagation channel $n=16$ in $RS$ region is close to 0.2 when $\sigma=0.1$, i.e. $<|V_{17,16}|^2>\sim 0.2$ shown in Fig. \ref{vans}(a), therefore the energy carried by the propagation channel $n=16$ will contribute nontrivially to the total transmission. While the value of $<|V_{17,17}|^2>$ is much larger than $<|V_{17,16}|^2>$, the energy transported by channel 17 in the $RS$ region is trivial since the channel $n=17$ in $RS$ region is evanescent, therefore the nontrivial transmission for incident channel 17, i.e. $<T_{17}>$ as shown in Fig. \ref{width}, is contributed from the coupling of energy from the incident channel 17 in $P_1$ region to the propagation channel 16 in $RS$ region. At the output interface between $RS$ and $P_2$ regions, a reverse coupling process occurs. 

When the disorder paramter $\sigma$ increases further, more eigen channels of the $RS$ region are roused and participate the process of carrying energy to the output interface, which subsequently increases the transmission for the incident channel $\alpha=17$, as shown in Fig. \ref{sigma}, since the value of the eigen channel of the $RS$ region are larger when the channel index is smaller, as shown in Fig. \ref{tau}. However, as shown in Fig \ref{vans}, the overall values of the coupling strength for channel $\alpha=17$ decrease when the disorder parameter $\sigma$ increases, therefore, the total transmission $T_\alpha$ for the incident channel $\alpha=17$ will reach a peak value and then decrease, as shown in Fig. \ref{sigma}, since $T_\alpha$ is determined by the weighting summation over the multiplication of coupling strength $|V_{\alpha n}|^2$ and eigenvalue of channels $\tau_n$.

Similar analysis can be performed for the cases of other incident evanescent channels, i.e. $\alpha=18,19$ here.

Therefore, by finely tuning the disorder strength of random scatterers in the $RS$ region, it is possible to realize an optimized transmission of light with a large incident angle, which is an evanescent mode in the corresponding homogeneous waveguide. Moreover, the transmission for light with a large incident angle can be possibly optimized further by artificially coupling into the propagation channels of the $RS$ region through the wavefront-shaping techniques, which is the underlying mechanism for the so-called scattering lens\cite{Vellekoop2007,Putten2011}.

\section{Conclusion}
In conclusion, we have used the RGF method and scattering-matrix method to investigate quantitatively in details the coupling of incident evanescent channels to the propagation channels in the random-scattering region. The energy carried by the incident evanescent channels, which were considered to be totally reflected at the interface, are coupled to the propagation channels in the random-scattering region, and then coupled to the output channels, which finally leads to the nontrivial energy transmission with a peak value for the incident evanescent channels. Our finding provides a clear and quantitative analysis on the far-field super-resolution imaging caused by evanescent-propagation transition mechanism, and will help to improve the far-field super-resolution imaging based on the wavefront-shaping technique.

\section*{Acknowledgement}
This work is supported by the National Natural Science Foundation of China under Grants No. 11374063, and the National Basic Research Program of China (973 Program) under Grant No. 2013CBA01505. Work at Ames Laboratory is partially supported by the U.S.Department of Energy, Office of Basic Energy Science, Division of Materials Science and Engineering (Ames Laboratory is operated for the U.S. Department of Energy by Iowa State University under Contract No. DE-AC02-07CH11358). The European Research Council under ERC Advanced Grant No. 320081 (PHOTOMETA) supports work at FORTH.

\section*{Reference}

\end{document}